\documentclass[aps,prb,preprint]{revtex4}

\input{tcilatex}
\usepackage{graphicx}

\begin{document}

\title{Symmetry of Two- and Four- Electron States in Solids. Application to
Unconventional Superconductors}
\author{V.G. Yarzhemsky}
\affiliation{Institute of General and Inorganic Chenistry of RAS, e-mail vgyar@igic.ras.ru}

\begin{abstract}
Group theoretical technique for construction of two-electron wavefunctions
with zero total momentum in solids based on the Mackey theorem on
symmetrized squares is developed. These states correspond to singlet and
triplet Cooper pairs. The nodal structure of these states is investigated
group theoretically and compared with experimental data for unconventional
superconductors. It is shown that when the Mackey theorem is applied twice,
the main four-electron states in solids can be constructed. Possible
connection of these states with experimental data is discussed.
\end{abstract}

\maketitle

\section{Introduction}

The antisymmetrical many-electron wavefunctions constitute the basis for
theory of strongly correlated electronic systems in atoms molecules and
solids \ The mathematical basis for such a technique in the spherically
symmetrical case is well developed (see e.g. [1,2]). On the other hand the
applications of permutational symmetry\ methods to solid state wavefunctions
are rather scant. The method of two-electron wavefunctions construction in
solids is mainly due to Bradley and Davies [3] who made use of Mackey
theorem on symmetrized squares of induced representations [4]. Theory of
relatively recently discovered unconventional superconductivity (i.e. high T$%
_{c}$ [5] and heavy fermion [6]) is not developed yet, and symmetry
considerations are very useful for understanding of electron structure of
these new materials. \ Non-totally symmetrical order parameter (or the
wavefunction of a Cooper pair) with line and point nodes on Fermi surface is
the mail feature of unconventional superconductors [7]. The induced
representation method \cite{b-d} was used for construction of
antisymmetrical wavefunctions in solids of different symmetry.[8-12]. This
group theoretical approach revealed a new symmetrical feature of
two-electron functions in solids : their nodal structure. Comparison of
group theoretical nodal structure for $D_{6h}$ group with experimental data
for unconventional superconductor $UPt_{3}$ resulted the $E_{2u}$ symmetry
for superconducting order parameter (SOP) [11]. The space group approach
makes possible construction of two-electron wavefunction by projection
operator technique [12]. But it is clear from the shape of these functions
that the commonly accepted interaction for a Cooper pair in $\vec{k}-\vec{k}$
manifold cannot give different energies for different irreducible
representations (IRs). Hence it follows that the interactions between other
wave vector are important in solids. Group theoretically it means that one
should construct the four electron wavefunctions with zero total momentum
i.e. Cooper quartets. In the present work the space-group approach to two
electron wavefunction is briefly reviewed and the theory is generalized to
four-electron wavefunctions. The classification of four electron states
connected by space inversion and mirror reflection is given making use Young
tables

\section{Tho-electron wavefunction in solids (Cooper pairs)}

From a unitary IR $t_{k}$ of a subgroup $H$ one can construct a unitary
representation of the whole group $G$ [13]. The structure of this unitary
representation (induced representation) depends on the left coset
decomposition of the whole group with respect to its subgroup: 
\begin{equation}
G=\sum_{i}s_{i}H  \label{1}
\end{equation}

where $i=1....n$ and $n=|G|/|H|$

The induced representation is defined by the following formula :

\begin{equation}
(t_{\kappa }\uparrow G)(g)_{i\mu ,j\nu }=t_{\kappa }(s_{i}^{-1}gs_{j})_{\mu
\nu }\delta (s_{i}^{-1}gs_{j},H)  \label{2}
\end{equation}

where:$\delta(s_{i}^{-1}gs_{j},H)={\Huge \{}%
\begin{array}{l}
1,\text{ if }\ s_{i}^{-1}gs_{j}\in H \\ 
0,\text{ if }s_{i}^{-1}gs_{j}\not \in H%
\end{array}
$

Following [13] we use an up directed arrow for the notation of induction.
The indexes $i$ and $j$ in formula denote the block columns and rows of the
induced representation matrix and correspond to the single coset
decomposition (1). The indexes $\mu $ and $\nu $ number the rows and columns
of the \textquotedblright small\textquotedblright\ IR \ $t_{k}$.

In the case of crystal symmetry the induced representation (2) is
irreducible representation of a space group, provided the group $H$ is a
wave vector $\vec{k}$ group ( \textit{little} \ group) and $t_{k}$ is its
unitary IR (\textit{small}\ IR). The action of left coset representatives $%
s_{i}$ on the wave vector $\vec{k}$ results all prongs of its star $\left\{ 
\vec{k}\right\} $. In the case strong spin-orbit coupling the IRs $t_{k}$ in
formula (3) are replaced by double-valued \textit{small} IRs $p_{k}$ .

According to the Pauli exclusion principle the total two-electron
wavefunction is antisymmetric with respect to permutation of electronic
coordinates. Hence in a weak spin-orbit coupling ($L-S$ scheme) the
symmetrized Kronecker square of the spatial part of the wavefunction is
combined with antisymmetrized Kronecker square of its spin part (singlet
pair), and the antisymmetrized Kronecker square of the spatial part of the
wavefunction is combined with the symmetrized Kronecker square of its spin
part (triplet pair). In a strong spin-orbit coupling case ($j-j$ scheme) the
wavefunction belongs to the antisymmetrized Kronecker square or
double-valued IR of the space group. According to Anderson [14] the Cooper
pair wavefunction is invariant with respect to lattice translations Hence it
follows the consideration is limited by the centre of a Brillouin zone for
two-electron states.

The structure of the Kronecker square of an induced IR may be envisaged by
the double coset decomposition of $G$ relative to $H$ which is written (3)
as:

\begin{equation}
G=\sum_{\sigma }Hd_{\sigma }H  \label{3}
\end{equation}

The sum runs over all distinct double cosets $\sigma.\ $

Corresponding wave vector $\overrightarrow{k}_{\sigma}$ is defined by the
following formula

\begin{equation}
\overrightarrow{k}+d_{\sigma }\overrightarrow{k}=\overrightarrow{k}_{\sigma
}+\overrightarrow{b}_{\sigma }  \label{4}
\end{equation}%
The intersection of wave vector groups in the left hand side is written as: 
\begin{equation}
M_{\sigma }=H\cap d_{\sigma }Hd_{\sigma }^{-1}  \label{5}
\end{equation}%
For each double coset we consider a representation of subgroup $M_{\sigma }$
defined by the formula:

\begin{equation}
P_{\sigma }=t_{\kappa }(m)\times t_{\kappa }(d_{\sigma }^{-1}md_{\sigma })
\label{6}
\end{equation}

Where $m\in M_{\sigma}$

For self-inverse double coset, i.e. : 
\begin{equation}
Hd_{\alpha }H=Hd_{\alpha }^{-1}H  \label{7}
\end{equation}

there are two extensions of \ $P_{\alpha }\ \ $into the subgroup: 
\begin{equation}
\tilde{M}_{\alpha }=M_{\alpha }+aM_{\alpha }  \label{8}
\end{equation}%
where$\ a=d_{\sigma }h_{1}=h_{2}d_{\sigma }\ $and$\ h_{1},h_{2}\in H$.

These extensions corresponding to symmetrized and antisymmetrized parts of
Kronecker square are defined in terms of their characters as follows:

\begin{equation}
\chi (P_{\alpha }^{+}(am))=+\chi (t_{\kappa }(amam))  \label{9}
\end{equation}

\begin{equation}
\chi (P_{\alpha }^{-}(am))=-\chi (t_{\kappa }(amam))  \label{10}
\end{equation}

$\ $where$\ m\in M_{\alpha}$.

The symmetrized and antisymmetrized parts of the Kronecker square of induced
representation are written by two following formulae respectively (the
Mackey theorem [3] on Kronecker squares):

\begin{equation}
\left[ t_{\kappa }\uparrow G\times t_{\kappa }\uparrow G\right] =\left[
q_{\kappa }\times q_{\kappa }\right] \uparrow G+\sum_{\alpha }P_{\alpha
}^{_{^{+}}}\uparrow G+\sum_{\beta }P_{\beta }\uparrow G  \label{11}
\end{equation}%
\begin{equation}
\left\{ t_{\kappa }\uparrow G\times t_{\kappa }\uparrow G\right\} =\left\{
t_{\kappa }\times t_{\kappa }\right\} \uparrow G+\sum_{\alpha }P_{\alpha
}^{_{^{-}}}\uparrow G+\sum_{\beta }P_{\beta }\uparrow G  \label{12}
\end{equation}

The first items on the right-hand sides of (11) and (12) correspond to the
double coset defined by the identity element, $\alpha $ corresponds to
self-inverse double cosets and $\beta $ to non-self-inverse double cosets
for which $Hd_{\beta }H\not=Hd_{\beta }^{-1}H$. \ In the case of a strong
spin-orbit coupling case the possible symmetries of two-electron states are
obtained by substituting of double-valued IRs into formula (12).

If the one-electron wave vector \ $\vec{k}$ belongs to a general point
inside a Brillouin zone the two-electron wave vector, defined by formula
(4), equals zero for the self-inverse double coset defined by the space
inversion. The extended intersection group $\tilde{M}_{\alpha }$ defined by
formula (8) is the group $C_{i}$ consisting of two elements : $E$ and $I$.
For single-valued IRs we immediately obtain that $P_{\alpha }^{_{^{+}}}$
equals to IR $A_{g}$ of group $C_{i}$ \ and that $P_{\alpha }^{_{^{-}}}$
equals to IR $A_{u}$ of group $C_{i}$.

These representations are induced into the central extension of the space
group (point group). The induced representation can be easily decomposed
making use of the Frobenius reciprocity theorem: the number of appearance of
\ the IR $\Gamma _{\kappa }$ of the whole group in the decomposition of the
induced representation $p_{k}\uparrow G$ equals to the number of appearance
of IR $p_{k}$.in the decomposition of $\Gamma _{\kappa }$when it subduced to
the subgroup. Making use of Frobenius theorem we obtain two conclusions.
Firstly, in agreement with Anderson (14) , we obtain that for $\vec{k}$ a
general point of a Brillouin zone all even IRs are possible for singlet
pairs and all odd IRs are possible for triplet pairs. Secondly, the number
of appearance of each IR equals to its dimension. Hence it follows that for
one-dimensional IR the result is unique, but for two-dimensional IRs there
are two non-equivalent basis functions and one can take any linear
combinations. From this point of view experimentally observed double
superconducting transition in $UPt_{3}$ (15) may be connected with two
non-equivalent states corresponding to the same two-dimensional IR . To
obtain total wavefunction of a Cooper pair in a weak spin-orbit coupling ($%
L-S$ scheme) one should multiply the spatial part of the wavefunction by
spin singlet function $S^{0}$ for singlet pair and by spin triplet function $%
S^{1}$ for triplet pair.

In strong spin orbital coupling case the representation $P_{\alpha
}^{_{^{-}}}$ equals to IR $A_{u}$ of group $C_{i}$ and even IRs are missing.
To obtain all possible pair symmetries, the time reversal $\theta $\ should
be considered. In the absence of magnetic fields the total symmetry of a
crystal with Fedorov group $G$ is described by the Shubnikov II (grey)
magnetic group: 
\begin{equation}
M=G+\theta G  \label{13}
\end{equation}

where $\theta$ is a time-reversal operation

The time-reversal symmetry results in additional degeneration for one-
dimensional small double-valued IRs i.e. at general points and at the planes
of symmetry in a Brillouin zone. To obtain all possible two-electron states
one should use induced corepresentations $D(p_{k})\uparrow G$ [13] in
formula (12).

For $\vec{k}$ a general point in a Brillouin zone the decomposition of
corepresentation $P_{\alpha }^{-}$ (see formula (6)) contains
representations $A_{g}$ and $3A_{u}$ of the group $C_{i}$. The IR $A_{g}$
corresponds to singlet pair and $3A_{u}$ correspond to three components of
triplet pair.

The superconducting state is usually more ordered than normal state, i.e.
the transition to it is accompanied by the symmetry reduction[16] . One
possible way is the time-reversal symmetry violation, i.e. transition from
the direct product $\theta \times G$ to ordinary Fedorov group $G$ or to one
the Shubnikov group $\theta \times (G-H)+H$ [13]. Total number of different
cases of construction of Shubnikov groups is quite large. In order to
envisage general trends we consider the simplified case of time reversal
symmetry violation and its influence on the nodal structure of
superconducting order parameter. Ferromagnetic fluctuations can be
approximated as time-reversal symmetry violation. In this case the
one-electron states belong to double valued IR of the space group. For $\vec{%
k}$ a general point of a Brillouin zone we obtain for two-electron states
two IRs $A_{u}$ of group : one for spin up states and one for spin-down
state. Two remaining IRs correspond to antiferromagnetic pairs: $A_{g}$ for
singlet pair and $A_{u}$ for triplet pair . Hence we obtain following
formula for the character of the possible Cooper pair representation
(reducible) in the antiferromagnetic state

\begin{equation}
\chi _{aitif}=\chi _{normal}-2\chi _{ferro}  \label{14}
\end{equation}%
Where $\chi _{normal}$ and $\chi _{ferro}$ are obtained respectively by
substitution of double valued \textit{small} \ corepresentation and
double-valued I into the formula antisymmetrized Kronecker square of induced
representation. It should be noted that formula (14) is valid in the case of
one-dimensional double-valued small IR for one-electron states, i.e. at
general points and at the planes of symmetry in one-electron Brillouin zone.

Making use of formula(14) for a general point in a Brillouin zone we obtain
that the Kronecker product decomposition for antiferromagnetic state
contains IRs $A_{g}$ and $A_{u}$ of the group $C_{i}$. Hence it follows that
in antiferromagnetic state even and odd Cooper pairs are possible. This
general result agrees with the experimental data which show both even and
odd symmetry for antiferromagnetic heavy-fermion superconductors $%
CeCu_{2}Si_{2}$ and $UPt_{3}$ respectively [7].

The space-group approach to the wavefunction of a Cooper pair makes it
possible to investigate the nodal structure of SOP as follows. One should
consider the distinct directions and planes of symmetry in a one-electron
Brillouin zone and calculate the antisymmetrized Kronecker squares with zero
total momenta of double-valued IRs or of double-valued corepresentations.
The absence of any IR in this square indicates a node of the SOP of this
symmetry. There are two types of nodes. The intersection of the direction of
nodes of any IR with the Fermi surface results in the point node. The
intersection of the plane of nodes with the Fermi surface results in the
line of nodes.

Possible IRs for all states at the plain of symmetry (group $C_{2h}$) are
presented in Table 1. In normal state all odd IRs are present and one even
IR $B_{g}$ of the group $C_{2h}$ is absent in the decomposition. Hence, it
follows that in this case, only nodes of even order parameter on the planes
of symmetry are required by the space-group symmetry and no limitations on
odd IRs exist. This statement is in agreement with the Blount [17]\cite%
{blunt} theorem according to which, \textit{it is vanishingly improbable\
for \textquotedblright triplet\textquotedblright\ superconductors to have
curves of vanishing gap on the Fermi surface}. If the time-reversal symmetry
is violated, the antisymmetrized square of the double-valued IR equals to IR 
$A_{u}$ of the group $C_{2h}$. Thus, in the ferromagnetic state, only odd
IRs are possible for the SOP on the planes of symmetry. The lack of the
second odd IR $B_{u}$ signifies that some of the odd IRs of the point group
are forbidden on the planes. The intersection of the plane with the Fermi
surface results in the line node of the odd SOP.

Going over to the antiferromagnetic state we see in Table 1 that one even IR 
$A_{g}$ (the same as in normal state) and one odd IR $B_{u}$ appear in the
decomposition. Hence it follows that the symmetry requirements for line
nodes of even IRs are the same as in normal state, but the lines of node of
odd IRs differ from that in ferromagnetic state.

Hence it follows that the theory is in agreement with the above mentioned
experimental on the SOP symmetry in unconventional superconductors, i.e.
antiferromagnetic superconductors may be either even (singlet) and odd
(triplet) with lines of nodes.

Another reason for violation of Blount theorem is due to crystal symmetry
lower then $O_{h}$.$T_{h}$ and $T_{d}$. In the case of $O_{h}$ symmetry spin
function belong to three dimensional IR $T_{1g}$. Following relation is
valid for the Kronecker product of the induced IR $\Gamma $ of the whole
group :

\begin{equation}
\Gamma \times (P_{\alpha }^{_{^{-}}}\uparrow G)=(\Gamma \downarrow \tilde{M}%
_{\alpha }\times P_{\alpha }^{_{^{-}}})\uparrow G  \label{15}
\end{equation}%
Hence we obtain that for $O_{h}$ symmetry all odd IRs of the subgroup $%
C_{2h} $ are possible for triplet pair and Blount theorem is fulfilled. For $%
D_{4h}$ and $D_{6h}$ symmetry the $M_{s}$=1 and -1 (or $S_{x}$ and $S_{y}$)
components belong to IR $E_{g}$ and $M_{s}$=0 (or $S_{z}$) belong to IR $%
A_{2g}$. It is natural to expect that due to interactions of spins with
crystal field the energies of spin states $E_{g}$ and $A_{2g}$ are different
and only one of them corresponds to superconducting state. Since not all IRs
of group $C_{2h}$ are present in the decomposition for both cases and the
lines of nodes appear. Thus another symmetry reason for violation of Blount
theorem is the lower crystal symmetry.

Table 1.

The decomposition of representations $P_{\alpha}^{-}$ and ${\small P}%
_{\alpha }^{_{^{+}}}$

\ for the planes of symmetry (group $C_{2h}$)

\begin{tabular}{|l|l|l|l|l|l|}
\hline
state & \multicolumn{4}{|l|}{character} & decomposition \\ \hline
& $E$ & $\sigma _{h}$ & $I$ & $C_{2}$ & IRs \\ \hline
normal & 4 & 0 & -2 & 2 & $A_{g}+2A_{u}+B_{u}$ \\ \hline
ferromagnetic & 1 & -1 & -1 & 1 & $A_{u}$ \\ \hline
antiferromagnetic & 2 & 2 & 0 & 0 & $A_{g}+B_{u}$ \\ \hline
${\small P}_{\alpha }^{_{^{+}}}$ & ${\small 1}$ & ${\small 1}$ & ${\small 1}$
& ${\small 1}$ & ${\small A}_{g}$ \\ \hline
${\small P}_{\alpha }^{_{^{-}}}$ & ${\small 1}$ & ${\small 1}$ & ${\small -1}
$ & ${\small -1}$ & ${\small B}_{u}$ \\ \hline
${\small P}_{\alpha }^{_{^{-}}}{\small \times T_{1g}\downarrow C}_{2h}$ & $%
{\small 3}$ & ${\small -1}$ & ${\small -3}$ & ${\small 1}$ & ${\small 2A}_{u}%
{\small +B}_{u}$ \\ \hline
${\small P}_{\alpha }^{_{^{-}}}{\small \times E}_{g}{\small \downarrow C}%
_{2h}$ & ${\small 2}$ & ${\small -2}$ & ${\small -2}$ & ${\small 2}$ & $%
{\small 2A}_{u}$ \\ \hline
${\small P}_{\alpha }^{_{^{-}}}{\small \times A}_{2g}{\small \downarrow C}%
_{2h}$ & ${\small 1}$ & ${\small 1}$ & ${\small -1}$ & ${\small -1}$ & $%
{\small B}_{u}$ \\ \hline
\end{tabular}

The wavefunction of \ a Cooper pair for the whole group may be obtained by
applying projection operators technique to the singlet and triplet
wavefunction in $\vec{k}-\vec{k}$ manifold. We will construct these
functions taking symmetry groups $D_{2h}$ and $D_{4h}$ of high temperature
superconductors. Let us denote $\vec{k}_{1}$ the wave vector chosen in the
representation domain of a Brillouin zone. Making use of Kovalev's [18]
notation $h_{25}$ for the space inversion the spatial parts of singlet and
triplet functions are written as:

\begin{equation}
\Phi _{1}^{s}=\psi _{1}^{1}\psi _{25}^{2}+\psi _{25}^{1}\psi _{1}^{2}
\label{16}
\end{equation}

\begin{equation}
\Phi _{1}^{t}=\psi _{1}^{1}\psi _{25}^{2}-\psi _{25}^{1}\psi _{1}^{2}
\label{17}
\end{equation}%
Where the superscript of $\psi $ denotes the number of electronic coordinate
and subscript of $\psi $ the prong of the $\vec{k}$-vector star.\medskip

Acting by $h_{2}$ (180$^{o}$ rotations around \ the axis $X$) on the
functions (16) and (17) we obtain two other basis functions:

\begin{equation}
\Phi _{2}^{s}=\psi _{2}^{1}\psi _{26}^{2}+\psi _{26}^{1}\psi _{1}^{2}
\label{18}
\end{equation}

\begin{equation}
\Phi _{2}^{t}=\psi _{2}^{1}\psi _{26}^{2}-\psi _{26}^{1}\psi _{2}^{2}
\label{19}
\end{equation}

Note that in Kovalev's \cite{koval} notations for $O_{h}$ group
multiplication of pure rotation element by $I$ corresponds to adding 24 to
the element number. To construct full basis for $D_{2h}$ group we need also
\ functions $\Phi _{3}^{s(t)}$ and $\Phi _{4}^{s(t)}$, which are obtained
from $\Phi _{1}^{s(t)}$\ by the action of 180$^{o}$ rotations around \ the
axes $Y$ and $Z$ respectively. In addition, for $D_{4h}$ group the elements $%
h_{13}$, $h_{16}$ (180$^{o}$ rotations around \ the \ axes $(\bar{1}10)$ and 
$(110)$) and $h_{14}$, $h_{15}$ $(90^{o}$ and $270^{o}$ counterclockwise
rotation around $Z$ axis) are required.

These functions span the space two-electron wavefunctions under the action
of all point group operations. Since the space inversion is already included
in the basis functions, their total number equals to the half of number of
point group operations. The action of pure rotations on the initial vector $%
\vec{k}_{1}$ result in a star, whose number of prongs is half of the number
of prongs in the the wave vector star. The action of the space inversion on
the basis vector corresponding to any prong doesn't change a vector but
introduces multiplier $-1$ for the triplet case. Making use of standard
projection operator technique and functions $\Phi _{1-4}^{s,t}$ we easily
obtain the basis functions for Cooper pairs belonging to all IRs of $D_{2h}$
group. The results are presented in Table 2.

Before going to the projection for $D_{4h}$ group it is useful to remind the
following correspondence of IRs in the subduction $D_{4h}$ $\downarrow
D_{2h} $ : $A_{1}$ \textit{and} $B_{1}\longrightarrow A_{1}$, $A_{2}$ 
\textit{and} $B_{2}$ $\longrightarrow B_{1}$, $E\longrightarrow B_{2}+B_{3}$%
. The basis functions for one-dimensional IRs of $D_{4h}$ group are
immediately obtained by projection operator technique. \ Since each of IRs $%
E_{g(u)}$ appear twice in the Kronecker product decomposition, there are two
independent basis sets labeled by additional quantum numbers . Bearing in
mind the above reduction scheme , we begin with basis sets corresponding to
IRs $B_{2}$ and $B_{3}$ of group $D_{2h}$ we obtain the remainder results of
Table 2, denoted by superscripts $\alpha $ and $\beta $ respectively. The
results for triplet pairs for $D_{4h}$ group pairs are not presented in the
Table 2. To obtain wavefunctions of triplet pairs one should replace
subscripts $g$ to $u $ in the first column and all superscripts $s$ to $t$
in the second column without changing of the signs.

For $\vec{k}$ a general point in a Brillouin zone all IRs are possible for
Cooper pair. But when the $\vec{k}$-vector approaches any mirror plane, the
mirror reflection image of $\vec{k}$ also approaches the $\vec{k}$-vector.
Total number of states decreases and lines of nodes are eventual. There are
two possibilities. If two-electron function is unchanged under the action of
the reflection, the function under consideration is nonvanishing on the
mirror plane. On the other hand, if the function changes its sign, two
mirror counterparts are cancelling on the plane. This corresponds to the
line of nodes. Note, that the space inversion changes the sign of the
spatial part of the triplet function. Making use of \ the above rules we can
easily obtain nodal structure of basis functions of one- dimensional IRs of
groups $D_{2h}$ and $D_{4h}$ presented in Table 2.

{\small TABLE 2. Spatial parts of Cooper pair wavefunctions for point groups 
$D_{2h}$ and $D_{4h}$}

\noindent 
\begin{tabular}{|l|l|l|l|}
\hline
\multicolumn{2}{|l|}{{\small D}$_{2h}$} & \multicolumn{2}{|l|}{{\small D}$%
_{4h}$} \\ \hline
{\small IR} & {\small pairing function} & {\small IR} & {\small pairing
function} \\ \hline
{\small A}$_{1g}$ & ${\small \Phi }_{1}^{s}{\small +\Phi }_{2}^{s}{\small %
+\Phi }_{3}^{s}{\small +\Phi }_{4}^{s}$ & {\small A}$_{1g}$ & ${\small \Phi }%
_{1}^{s}{\small +\Phi }_{2}^{s}{\small +\Phi }_{3}^{s}{\small +\Phi }_{4}^{s}%
{\small +\Phi }_{13}^{s}{\small +\Phi }_{14}^{s}{\small +\Phi }_{15}^{s}%
{\small +\Phi }_{16}^{s}$ \\ \hline
{\small B}$_{1g}$ & ${\small \Phi }_{1}^{s}{\small -\Phi }_{2}^{s}{\small %
-\Phi }_{3}^{s}{\small +\Phi }_{4}^{s}$ & {\small A}$_{2g}$ & ${\small \Phi }%
_{1}^{s}{\small -\Phi }_{2}^{s}{\small -\Phi }_{3}^{s}{\small +\Phi }_{4}^{s}%
{\small -\Phi }_{13}^{s}{\small +\Phi }_{14}^{s}{\small +\Phi }_{15}^{s}%
{\small -\Phi }_{16}^{s}$ \\ \hline
{\small B}$_{2g}$ & ${\small \Phi }_{1}^{s}{\small -\Phi }_{2}^{s}{\small %
+\Phi }_{3}^{s}{\small -\Phi }_{4}^{s}$ & {\small B}$_{1g}$ & ${\small \Phi }%
_{1}^{s}{\small +\Phi }_{2}^{s}{\small +\Phi }_{3}^{s}{\small +\Phi }_{4}^{s}%
{\small -\Phi }_{13}^{s}{\small -\Phi }_{14}^{s}{\small -\Phi }_{15}^{s}%
{\small -\Phi }_{16}^{s}$ \\ \hline
{\small B}$_{3g}$ & ${\small \Phi }_{1}^{s}{\small +\Phi }_{2}^{s}{\small %
-\Phi }_{3}^{s}{\small -\Phi }_{4}^{s}$ & {\small B}$_{2g}$ & ${\small \Phi }%
_{1}^{s}{\small -\Phi }_{2}^{s}{\small -\Phi }_{3}^{s}{\small +\Phi }_{4}^{s}%
{\small +\Phi }_{13}^{s}{\small -\Phi }_{14}^{s}{\small -\Phi }_{15}^{s}%
{\small +\Phi }_{16}^{s}$ \\ \hline
{\small A}$_{1u}$ & ${\small \Phi }_{1}^{t}{\small +\Phi }_{{\small 2}}^{t}%
{\small +\Phi }_{3}^{t}{\small +\Phi }_{4}^{t}$ & {\small E}$_{g}^{\alpha }$
& ${\small \Phi }_{13}^{s}{\small -\Phi }_{15}^{s}{\small +\Phi }_{14}^{s}%
{\small -\Phi }_{16}^{s}$ \\ \hline
{\small B}$_{1u}$ & ${\small \Phi }_{1}^{s}{\small -\Phi }_{2}^{s}{\small %
-\Phi }_{3}^{s}{\small +\Phi }_{4}^{s}$ &  & ${\small \Phi }_{1}^{s}{\small %
-\Phi }_{2}^{s}{\small +\Phi }_{3}^{s}{\small -\Phi }_{4}^{s}$ \\ \hline
{\small B}$_{2u}$ & ${\small \Phi }_{1}^{s}{\small -\Phi }_{2}^{s}{\small %
+\Phi }_{3}^{s}{\small -\Phi }_{4}^{s}$ & {\small E}$_{g}^{\beta }$ & $%
{\small \Phi }_{1}^{s}{\small +\Phi }_{2}^{s}{\small -\Phi }_{3}^{s}{\small %
-\Phi }_{4}^{s}$ \\ \hline
{\small B}$_{3u}$ & ${\small \Phi }_{1}^{s}{\small +\Phi }_{2}^{s}{\small %
-\Phi }_{3}^{s}{\small -\Phi }_{4}^{s}$ &  & ${\small \Phi }_{13}^{s}{\small %
+\Phi }_{15}^{s}{\small -\Phi }_{14}^{s}{\small -\Phi }_{16}^{s}$ \\ \hline
\end{tabular}

Two dimensional IRs appear twice for $\vec{k}$ a general point in a
Brillouin zone. In this case a direct analysis of nodal structure of basis
functions of Table 2 is required. The analysis shows that basis functions of
IR $E_{g}^{\alpha }$ vanish in the planes$(100)$ and $(001)$ and that of IR $%
E_{g}^{\beta }$ vanish in the planes $(010)$ and $(001).$ Linear
combinations of these basis functions $E_{g}^{\alpha }\pm E_{g}^{\beta }$
vanish in planes $(110)$ and $(\bar{1}10)$ respectively and both vanish in
plane $(001)$. Hence it follows that only lines of nodes in basal plane \
follow unambiguously from the symmetry. It should be noted, that point
group\ approach also results different nodal structure of different
two-dimensional IRs [19] .

The analysis of broad set of experimental data on the of high-T$_{c}$
superconductors [20] led the most of the authors to the conclusion of
singlet pairing and $A_{g}$ SOP symmetry in these compounds. Angular
resolved photoelectron spectra of high-T$_{c}$ superconductors reveal a
strong trough in the diagonal of $xy$ plane indicating $d_{x^{2}-y^{2}}$
-pairing with line of nodes. On the other hand some experiments reveal also
totally symmetric $s$ -pairing without nodes. In many cases an interplay
between these two types of pairing [20] both belonging to $A_{g}$ IR exists.
It is seen from Table 2, that $A_{g}$ pairing function, obtained group
theoretically is noddles and that other IRs have nodes in the coordinate
planes only. Hence it follows that nodal structure of high-T$_{c}$
superconductors is more complex then that which follows from the symmetry
only. To explain this one can consider two wave vectors $\vec{k}_{\alpha }$
and $\vec{k}_{\beta }$ symmetrical with respect to diagonal of the deformed
square. Note, that the orthorombicity $[(b-a)/(b+a)]$ of $YBCuO$ is about 2
\% only [20]. Two types of basis functions of Cooper pairs belonging to $%
A_{g}$ IR $\Phi _{\alpha }^{s}$ and $\Phi _{\beta }^{s}$ are easily obtained
from the Table 2 by introducing additional subscripts $\alpha $ and $\beta $%
. One can suppose that due to the interaction two self-vectors are linear
combinations of these basis states :

\begin{equation}
\Phi _{1}^{S}=C_{\alpha }\left( \Phi _{1,\alpha }^{S}+\Phi _{2,\alpha
}^{S}+\Phi _{3,\alpha }^{S}+\Phi _{4,\alpha }^{S}\right) +C_{\beta }\left(
\Phi _{1,\beta }^{S}+\Phi _{2,\beta }^{S}+\Phi _{3,\beta }^{S}+\Phi
_{4,\beta }^{S}\right)  \label{20}
\end{equation}

\begin{equation}
\Phi _{2}^{S}=C_{\beta }\left( \Phi _{1,\alpha }^{S}+\Phi _{2,\alpha
}^{S}+\Phi _{3,\alpha }^{S}+\Phi _{4,\alpha }^{S}\right) -C_{\alpha }\left(
\Phi _{1,\beta }^{S}+\Phi _{2,\beta }^{S}+\Phi _{3,\beta }^{S}+\Phi
_{4,\beta }^{S}\right)  \label{21}
\end{equation}

Both combinations belong to IR $A_{g}$ of group $D_{2h}$. First one
corresponds to the noddles $s$-pairing and the second to the $%
d_{x^{2}-y^{2}}-$pairing with line of nodes in the diagonal of $xy$- plane.
In the limit of zero orthorombic distortion the symmetry group is $D_{4h}$
and $C_{\alpha }=C_{\beta }$ , subscripts \ $\alpha $ and $\beta $ are
dropped and the sums\ in second brackets in right hand sides of (20) and
(21) are written as $\Phi _{13}^{S}+\Phi _{14}^{S}+\Phi _{15}^{S}+\Phi
_{16}^{S}$. In this case combination (20) belongs to IR $A_{1g}$ and
combination (21) belongs to IR $B_{1g}$ of the symmetry group $D_{4h}$.
Hence it follows that the nodal structure of SOP in high- T$_{c}$
superconductors is defined by hidden symmetry $D_{4h}.$

\section{Four-electron wavefunctions}

It is clear from the shape of two-electron wavefunctions presented in Table
2 that if the interaction within the $\vec{k}-\vec{k}$ manifold only, is
anticipated, the energies of all IRs of the same parity are the same. This
statement contradicts with the experimental fact that the superconductivity
in unconventional superconductors is defined by only one IR of crystal point
group. Hence it follows that we must suppose additional residual
interaction, say between electrons connected by mirror reflection operation $%
\sigma _{h}$. In this case one should consider four electron state connected
by space inversion, 180$^{o}$ rotation and mirror reflection. We consider
the the four -electron states which may be constructed on the basis of the
singlet and triplet pairs in $\vec{k}-\vec{k}$ manifold, whose spatial part
correspond to Young tables $[2]$ and $[1^{2}]$ respectively. In general case
it is done by a plethism operation [1]. In our case the induced structure of
IRs makes possible using of the Mackey theorem as follows. The spatial parts
of all singlet pairs are given by second term in the right hand side of
(11). All four electron states constructed from singlet pairs are contained
in the Kronecker product $(P_{\alpha }^{_{^{+}}}\uparrow G)\times (P_{\alpha
}^{_{^{+}}}\uparrow G)$. Due the induced structure of these representations
one can use the Mackey theorem ones more, taking any element except the
space inversion as a coset representative. In this second step the
symmetrized square of singlet spatial part is excluded, since it results the
Young table $[4]$ whose conjugate spin part $[1^{4}]$ is forbidden. Taking
180$^{o}$ rotation around axis perpendicular to the plane and mirror
reflection as a coset representatives in double coset decomposition (3) we
obtain all four electron states shown in figures 1 a) and 1 b) respectively.

\FRAME{ftbpF}{4.0782in}{2.5119in}{0pt}{}{}{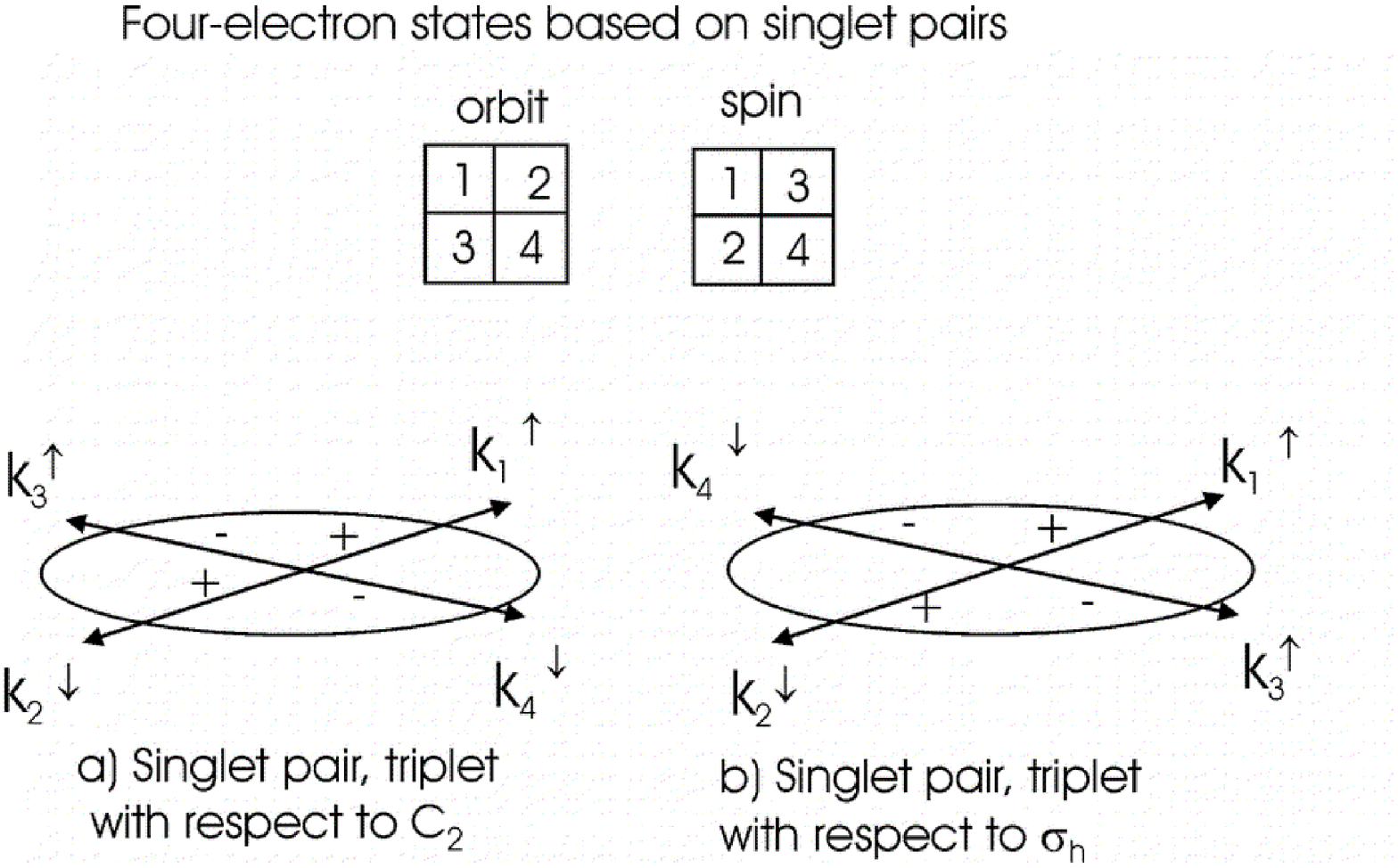}{\raisebox{-2.5119in}{\includegraphics[height=2.5119in]{fig1.ps}}}In this figure the parity with respect to $\vec{k}$ is indicated
by the sign and the spin direction by an arrow. \ Note that the Young tables
corresponding to both pictures are the same, but indexes 3 and 4 are
interchanged in pictures. All these states are odd with respect to mirror
reflection, but no limitations are imposed on the spin orientation of mirror
counterparts.

Similar procedure can be applied to triplet pairs. In this case in the
second step of application of Mackey theorem one can make both
symmetrization and antisymmetrization. But in the case of totally
antisymmetric spatial part one obtains a four electron state with totally
symmetrical spin part, i.e. with parallel spins. Such a particle with total
spin equals to 2 is not appropriate candidate for superconductivity and is
excluded from the present consideration.

The four electron states constructed from triplet pairs by symmetrization
with respect to double coset defined by 180$^{o}$ rotation and mirror
reflection are shown in figures \ 2 a) and 2 b) respectively. These states
belong to IRs $A_{u}$ and $B_{u}$ of the group $\sigma _{h}$ can have
different parity with respect to mirror reflection. The electrons connected
by mirror reflection have different spin directions in both cases.

\FRAME{ftbpF}{4.6372in}{2.5119in}{0pt}{}{}{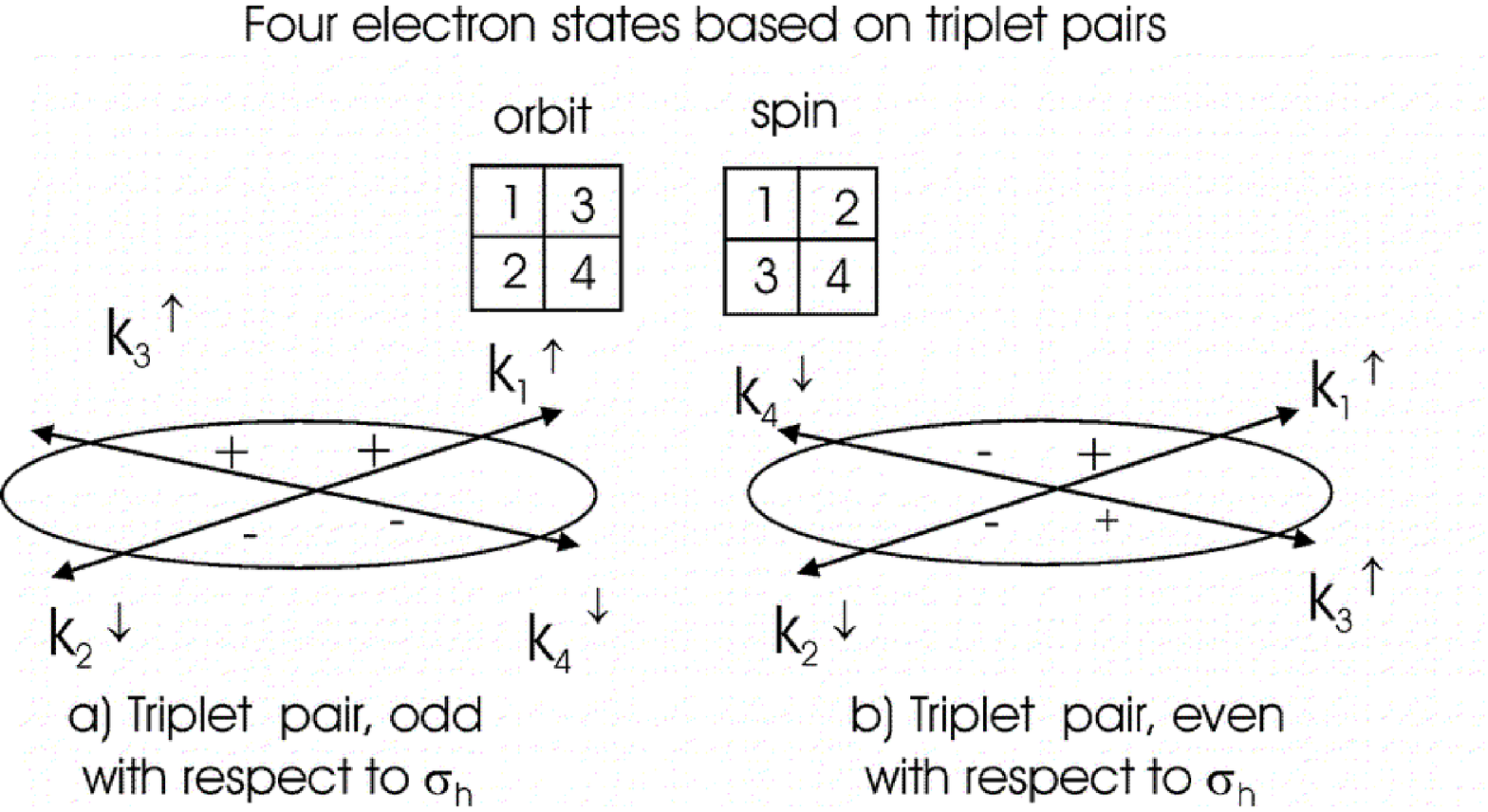}{\raisebox{-2.5119in}{\includegraphics[height=2.5119in]{fig2.ps}}}The projection operator technique for two-electron states
developed in the previous section results all even IRs for singlet states
and all odd IRs for triplet case. When the four-electron states are
constructed, the singlet states, even with respect to 180$^{o}$ rotation are
excluded. It should be noted that additional degeneration due to time
reversal symmetry should be taken into account for two-electron states and
we can apply or not apply time-reversal in addition to mirror reflection. It
is seen from Figure 2 that in "triplet" four-electron case we have only one
possibility for spin orientation connected by mirror reflection. Hence it
follows that the total number of IRs possible for four-electron
quasiparticles is less then that in the case of two-electron quasiparticle .

\section{Conclusion}

The method for construction of two- and four- electron wavefunction in
solids based on Mackey theorem on symmetrized squares of induced
representations is developed. The nodal structure of two-electron states is
investigated group theoretically. It is shown that lines of nodes always
exist in a singlet case and appear in a triplet case if several symmetry
operations are violated. It is shown that the main four-electron states
(related to both singlet and triplet pairs in $\vec{k}-\vec{k}$ manifold)
are described by Young table $[2,2]$. Two advantages for introducing of
four-electron states into the theory of superconductivity are emphasized.
Firstly they correspond to interactions beyond $\vec{k}-\vec{k}$ manifold
and are able to explain different energies for different singlet and triplet
IRs. Secondly total number or permitted symmetries for SOP is reduced.

\section{References}

1. B.G.Wybourne, \emph{Symmetry Principles in Atomic Spectroscopy, }Mir,
Moscow 1973.

2. B.R. Judd, \emph{Operator Techniques} \emph{in Atomic Spectroscopy,}, New
York 1963.

3. C.J. Bradley and B.L. Davies, J. Math. Phys.\textbf{\ 11} 1536 (1970).

4. G.W.Mackey, Am.J.Math. \textbf{75}, 387 (1953).

5. J.G. Bednorz and K.A. M\"{u}ller: Z. Phys. B \textbf{64} 189(1986) .

6. D.J. Bishop, C.M. Varma, B. Batlogg, E. Bucher, Z. Fisk and J.L. Smith:
Phys. Rev. Lett \textbf{53} 1009 (1984).

7. M.Sigrist and K.Ueda, Rev. Mod.Phys. \textbf{63}, 239 (1991)

8. V.G. Yarzhemsky and E.N. Murav'ev: J. Phys: Cond. Matter 4 (1992) 3525.

9. V.G. Yarzhemsky: Zeitsch. Phys. B 99 (1995) 19.

10. V.G.Yarzhemsky: Phys. stat. sol. (b) 209 (1998) 101.

11. V.G. Yarzhemsky: Int. J. Quant. Chem. Int. J. Quant. Chem. 80 (2000) 133.

12. V.G.Yarzhemsky: in Lectures on the Physics of Higly Correlated Electron
Systems, p.343--252, Melille, N.Y.2003.

13. C.J. Bradley and A.P. Cracknell: \emph{The Mathematical Theory of
Symmetry in Solids. Representation Theory of Point Groups and Space Groups.}
Oxford, Clarendon, 1972.

14. P.W. Anderson: Phys. Rev. B 30 (1984) 4000.

15. R.A.Fisher, S.Kim, B.F.Woodfield, Phys.Rev.Lett., 62, 1411 (1989).

16 V.V.Schmidt, \emph{Intruduction into Superconductors Physics. }Moscow,
MCNMO 2000 (in Russian)

17. E.I. Blount: Phys. Rev. B 32 (1985) 2935.

18. O.V. Kovalev:\emph{\ Irreducible and Induced Representations and
Corepresentations of Fedorov Groups }Nauka 1986, (in Russian).

19 G.E. Volovik and L.P. Gor'kov: Sov. Phys. Jetp. 61 (1985) 843.

2. C.C. Tsuel and J.R. Kirtley: Rev. Mod. Phys. 72 (2000) 969.

\end{document}